\newcommand{\tr}{\rm tr \,}
\documentclass[final]{aipproc}
\layoutstyle{6x9}
\usepackage{epsfig}
\begin{document}

\title{Charmed meson resonances from\\  chiral coupled-channel dynamics}
\author{E.E. Kolomeitsev}{address={The Niels Bohr Institute\\ Blegdamsvej 17,
DK-2100 Copenhagen, Denmark} }
\author{M.F.M. Lutz}{address={Gesellschaft f\"ur Schwerionenforschung (GSI)\\
Planck Str. 1, D-64291 Darmstadt, Germany} }

\begin{abstract}
Charmed meson resonances with quantum numbers $J^P=0^+$ and
$J^P=1^+$ are generated in terms of chiral coupled-channel
dynamics. At leading order in the chiral expansion a
parameter-free prediction is obtained for the scattering of
Goldstone bosons off charmed pseudo-scalar and vector mesons.
The recently announced narrow open charm states observed by the
BABAR and CLEO collaborations are reproduced. We suggest the
existence of states that form an anti-triplet and a sextet
representation of the SU(3) group. In particular, so far
unobserved narrow isospin-singlet states with negative strangeness
are predicted.
\end{abstract}

\maketitle


Recently a new narrow state of mass 2.317~GeV that decays into
$D^+_s\,\pi^0$ was announced \cite{BaBar}. This result was
confirmed \cite{CLEO} and a second narrow state of mass 2.463~GeV
decaying into $D_s^* \pi^0$ was observed. Such states were first
predicted in \cite{NRZ93,BH94} based on the spontaneous breaking
of chiral symmetry. The theoretical predictions
\cite{NRZ93,BH94,BEH03,NRZ03} rely on the chiral quark model which
predicts the heavy-light $0^+,1^+$ resonance states to form an
anti-triplet representation of the SU(3) group. If one insists on
a non-linear realization of the chiral SU(3) group and excludes
any further model assumptions, no a priori prediction can be made
for the existence of chiral partners of any given state.

Recently it was shown \cite{LK03} that solving the coupled-channel
Bethe-Salpeter equation with the interaction kernel following from
a non-linear chiral SU(3) Lagrangian one is able to predict two
octets and a singlet multiplets of the light $1^+$ mesons
consistent with the empirical spectrum.
Similar results were obtained for light meson resonances
with $J^P=0^+$ quantum numbers
\cite{Rupp86,WI90,JPHS95,OOP99,NVA02,NP02}.
In view of the evident success of the chiral coupled-channel
dynamics to predict the existence of a wealth of meson and baryon
resonances in the ($u,d,s$)-sector of QCD
\cite{LK03,Granada,Copenhagen} it is expectable that the chiral
SU(3) symmetry should  also predict spectra of hadrons with open charm or beauty (see
also \cite{BR03}).
In this talk we review the $\chi$-BS(3)
approach \cite{LK00,LK01,LK02,LH02,LK03,Granada,Copenhagen} as
applied to open-charm meson resonances \cite{KL03,HL03}.
We will not have space to discuss further exciting results concerning
open-beauty meson resonances \cite{KL03}, open-charm baryon
resonances \cite{LK03-baryon} or results in the ($u,d,s$)-sector
of QCD \cite{LK03,Granada,Copenhagen}.


Heavy-light meson states with quantum numbers $J^P=0^+$ and
$J^P=1^+$ may be studied by considering the s-wave scattering of
Goldstone bosons off the heavy-light ground state mesons with
$J^P=0^-$ and $J^P=1^-$. If scalar or axial vector resonances
exist they should manifest themselves as poles in the
corresponding scattering amplitudes. The starting point to
describe low-energy scattering processes is the chiral SU(3)
Lagrangian \cite{Wein-Tomo,Wise92,YCCLLY92,BD92} including
heavy-light $0^-$ and $1^-$ fields. A systematic approximation
scheme arises due to a successful scale separation justifying
chiral power counting rules \cite{book:Weinberg}. Our effective
field theory for the scattering of Goldstone bosons off any heavy
field is based on the assumption that the scattering amplitudes
are perturbative at subthreshold energies with the expansion
parameter $Q/ \Lambda_{\chi}$. The small scale $Q$ is to be
identified with any small momentum of the system. The chiral
symmetry-breaking scale is $$\Lambda_\chi \simeq 4\pi f \simeq
1.13~{\rm GeV}\,, $$ with the parameter $f\simeq 90$ MeV
determined by the pion decay process. Once the available energy is
sufficiently high to permit elastic two-body scattering a further
typical dimensionless parameter $m_K^2/(8\,\pi f^2) \sim 1$ arises
\cite{LK00,LK01,LK02} if strangeness is considered explicitly.
This extra parameter invalidates any perturbative calculation
within chiral SU(3) effective theory. Since this ratio is uniquely
linked to two-particle reducible diagrams it is sufficient to sum
those diagrams keeping the perturbative expansion of all
irreducible diagrams, i.e. the coupled-channel Bethe-Salpeter
equation has to be solved. This is the basis of the $\chi$-BS(3)
approach developed in \cite{LK00,LK01,LK02,LK03}.

We identify the leading-order Lagrangian
density \cite{Wein-Tomo,Wise92,YCCLLY92,BD92} describing the interaction of
Goldstone bosons with pseudo-scalar and vector mesons,
\begin{eqnarray}
{\mathcal L}(x) &=&
\frac{1}{8\,f^2}\,\tr  \Big[
P_{\phantom{\mu}}(x)\, (\partial^\nu P^\dagger_{\phantom{\mu}}(x) )
-(\partial^\nu P_{\phantom{\mu}}(x))\,P^\dagger_{\phantom{\mu}}(x)  \Big]
\,[\Phi (x) , (\partial_\nu\,\Phi(x))]_-
\nonumber\\
&-& \frac{1}{8\,f^2}\,\tr
\Big[
P^\mu(x)\, (\partial^\nu P^\dagger_\mu(x) )-(\partial^\nu P^\mu(x))\,P^\dagger_\mu(x)  \Big]
\,\Big[\Phi (x) , (\partial_\nu\,\Phi(x))\Big]_- \,,
\label{WT-term}
\end{eqnarray}
where $\Phi$ is the octet of Goldstone boson fields  and $P$
and $P_\mu$ are the triplets of massive pseudo-scalar and vector-meson fields in the matrix
representation.

Within the $\chi-$BS(3) approach \cite{LK02,LK03} the s-wave scattering amplitudes,
$M^{(I,S)}_{J^P}(\sqrt{s}\,)$ take the simple form
\begin{eqnarray}
&&  M^{(I,S)}_{J^P}(\sqrt{s}\,) = \Big[ 1- V^{(I,S)}(\sqrt{s}\,)\,J^{(I,S)}_{J^P}(\sqrt{s}\,)\Big]^{-1}\,
V^{(I,S)}(\sqrt{s}\,)\,.
\label{final-t}
\end{eqnarray}
The effective interaction kernel $V^{(I,S)}(\sqrt{s}\,)$ in (\ref{final-t}) is determined by
the leading order chiral SU(3) Lagrangian (\ref{WT-term}),
\begin{eqnarray}
V^{(I,S)}(\sqrt{s}\,) = \frac{C^{(I,S)}}{8\,f^2}\, \Big(
3\,s-M^2-\bar M^2-m^2-\bar m^2
 -\frac{M^2-m^2}{s}\,(\bar M^2-\bar m^2)\Big) \,,
\label{VWT}
\end{eqnarray}
where $(m,M)$ and $(\bar m, \bar M)$ are the masses of initial and final mesons. We use
capital $M$ for the masses of heavy-light mesons and small $m$ for the masses of the Goldstone
bosons. The matrix of coefficients $C^{(I,S)}$, that characterize the interaction strength in
a given channel, and the loop functions $J^{(I,S)}_{J^P}(\sqrt{s}\,)$ are given \cite{LK03}.
As expected from heavy-quark symmetry the interaction kernels as well as the loop functions
are identical for the $0^-$ and $1^-$ sectors in the limit $M\to \infty$.

In order to guarantee the perturbative nature of the scattering
amplitude at subthreshold energies the $\chi-$BS(3) approach
insists on a renormalization condition of the form
\begin{eqnarray}
M^{(I,S)}(\sqrt{s}=\mu^{(I,S)}) = V^{(I,S)}(\sqrt{s}=\mu^{(I,S)})
\label{ren-cond}
\end{eqnarray}
with the natural subtraction scales
\begin{eqnarray}
&& \mu_{0^+}^{(I,0)} =  M_{D(1867)}\,,\quad
\mu_{0^+}^{(I, \pm 1)} = M_{D_s(1969)}\,,\quad
\mu_{0^+}^{(I, 2)} = M_{D(1867)} \,,
\nonumber\\
&& \mu_{1^+}^{(I,0)} =  M_{D(2008)}\,,\quad
\mu_{1^+}^{(I, \pm 1)} = M_{D_s(2110)}\,,\quad
\mu_{1^+}^{(I, 2)} = M_{D(2008)} \,.
\label{mu-def}
\end{eqnarray}
A crucial ingredient of the $\chi$-BS(3) approach is a matching of
s- and u-channel unitarized scattering amplitudes at subthreshold energies \cite{LK02,LK03}.
This construction reflects our basic assumption that diagrams showing an s-channel
or u-channel unitarity cut need to be summed to all orders at least at energies close to
where the diagrams develop their imaginary part. By construction, a matched scattering amplitude
satisfies crossing symmetry exactly at energies where the scattering process takes
place. At subthreshold energies crossing symmetry is implemented
approximatively only, however, to higher and higher accuracy when more chiral correction
terms are considered. Insisting on the renormalization condition
(\ref{ren-cond},\ref{mu-def}) guarantees that subthreshold amplitudes match smoothly
and therefore the final 'matched' amplitudes comply with the crossing-symmetry constraint
to high accuracy. A conceivable small variation of the subtraction scales around their natural
values (\ref{mu-def}) has very little effect on the results. In fact chiral correction terms
modify the effective interaction $V(\sqrt{s})$ rather than giving rise to a modification of the
subtraction scale \cite{LK02,LK03,HL03}. Changing the optimal subtraction scale (\ref{mu-def})
would deteriorate the quality of the matching of u- and s-channel unitarized amplitudes
\cite{LK02,LK03}.

\begin{figure}
\includegraphics[width=.47\textwidth]{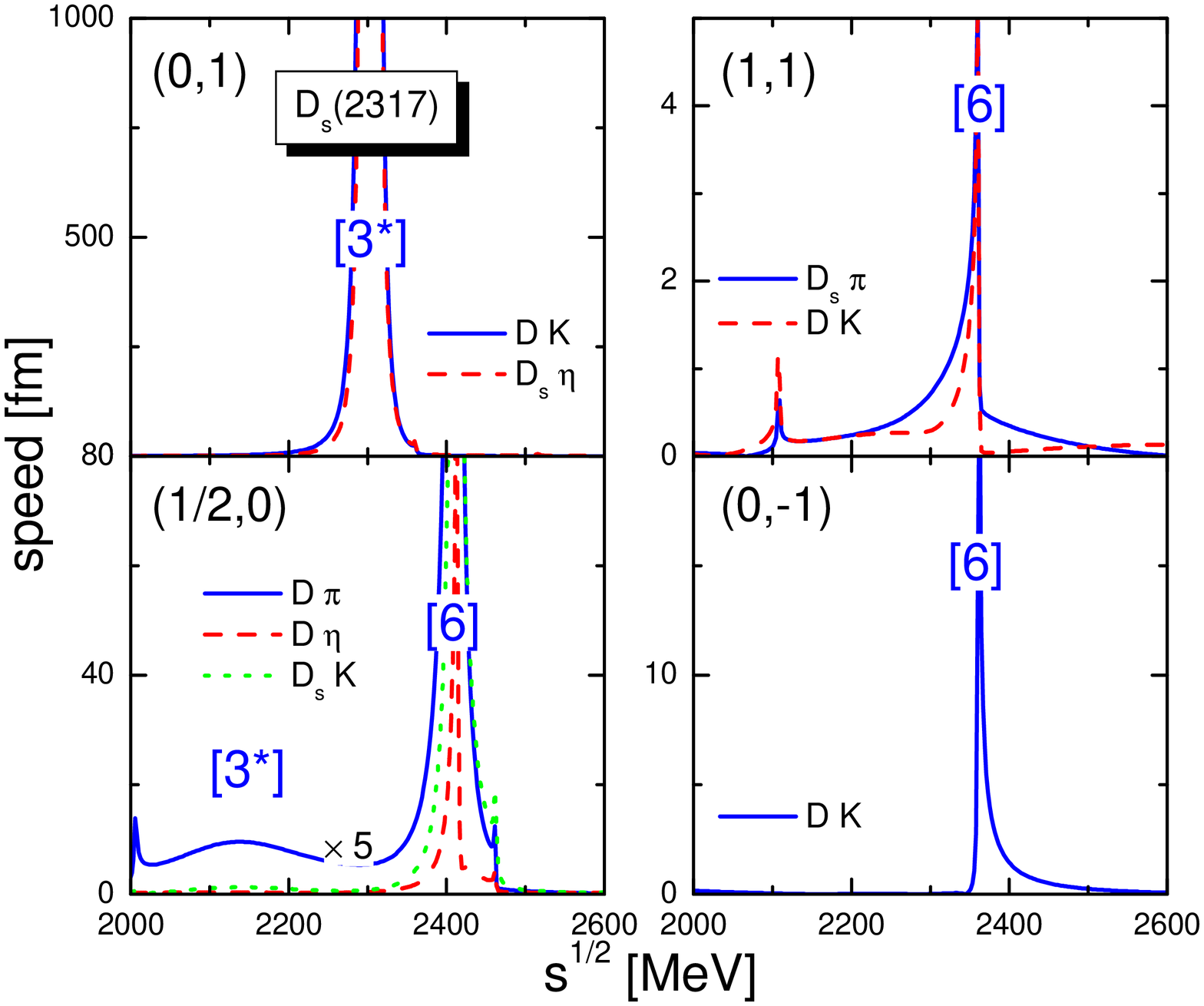}\hfill
\includegraphics[width=.47\textwidth]{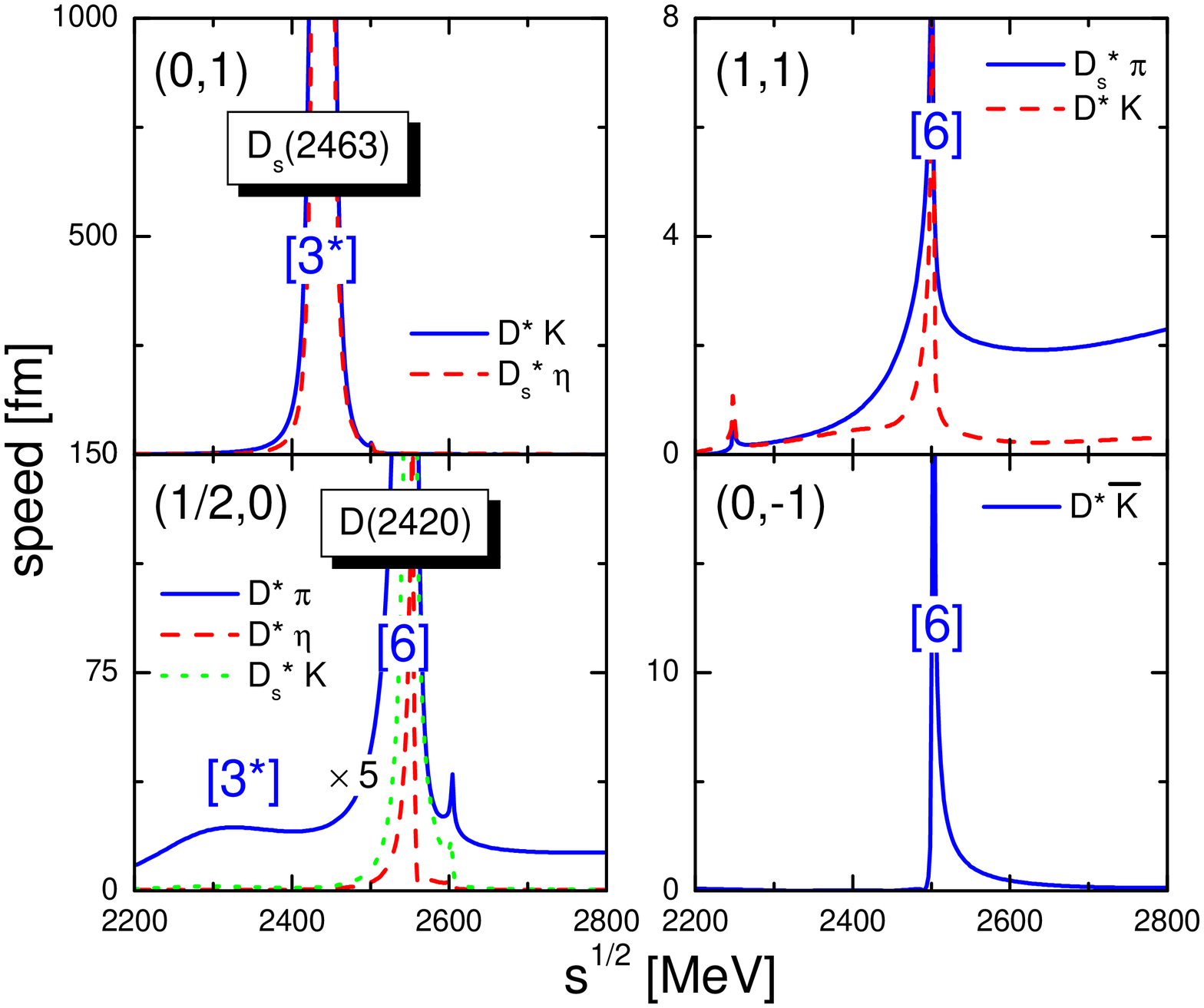}
\caption{Speed plots for heavy-light scalar (left panel) and axial-vector (right panel)
mesons with isospin ($I$) and
strangeness ($S$).}\label{fig:1}
\end{figure}

We turn to the results of the $\chi$-BS(3) approach for charmed
mesons. It is instructive to explore first the SU(3) multiplet
structure of  the resonance states formed by the chiral
coupled-channel dynamics. First the $0^+$ sector is discussed in a
'heavy' SU(3) limit \cite{Copenhagen,LK03} with $m_{\pi ,K,\eta} =
500$ MeV and $M_{D}=1800$ MeV. In this case we obtain an
anti-triplet of mass 2204 MeV with poles in the $(0,+1),(1/2,0)$
amplitudes. The sextet channel does not show a bound-state signal
in this case. However if the attraction is increased slightly by
using $f= 80$ MeV rather than the canonical value $90$ MeV, poles
at mass 2298 MeV arise in the $(1,+1),(1/2,0),(0,-1)$ amplitudes.
This finding reflects that the Weinberg-Tomozawa interaction,
\begin{eqnarray}
\bar 3\otimes 8= \bar 3\oplus 6 \oplus \overline{15}
\end{eqnarray}
predicts attraction in the anti-triplet and sextet channel but
repulsion for the anti-15-plet. In contrast performing a 'light'
SU(3) limit \cite{Copenhagen,LK03} with $m_{\pi,K,\eta} \sim
140$~MeV together with $M_{D}=1800$~MeV we do not find any signal
of a resonance in both anti-triplet and sextet channels. Analogous
results are found in the $1^+$ sector.

To study the formation of meson resonances  we generate speed
plots~\cite{speed} as suggested by H\"ohler~\cite{Hoehler:speed}
(for definitions cf.~\cite{KL03}). Fig.~\ref{fig:1} shows the
spectra of $0^+$ (left panel) and $1^+$ (right panel) as they
arise in calculations with physical masses. We predict a bound
state of mass 2303~MeV in the $(0,1)$-sector of $0^+$ mesons (see
Fig.~\ref{fig:1}, left panel). According to \cite{BEH03,NRZ03}
this state can be identified with a narrow resonance of mass
2317~MeV recently observed by the BABAR collaboration
\cite{BaBar}. Since we do not consider isospin violating processes
like $\eta \to \pi_0$ the latter state is a true bound state in
our present scheme. Given the fact that our computation is
parameter-free this is a remarkable result. In the $(1,+1)$-speeds where we
expect a signal from  the sextet a strong cusp effect at the
$K\,D(1867)$-threshold is seen. The large coupling constant to the
$\pi\,D_s(1969)$ channel leads to the broad structure seen in the
figure. Fig.~\ref{fig:1} (left panel) illustrates that in the
$(\frac{1}{2},0)$-sector we predict a narrow $0^+$ state of mass 2413
MeV just below the $\eta\,D(1867)$-threshold and a broad state of
mass 2138 MeV. Modulo some mixing effects the heavier of the two
is part of the sextet the lighter a member of the anti-triplet.
The latter $(\frac{1}{2},0)$-state was expected to have a large
branching ratio into the $\pi \,D(1867)$-channel
\cite{BR03,NRZ03}. This is confirmed by our analysis. Finally in
the $(0,-1)$-speed a pronounced cusp effect at the $\bar
K\,D(1867)$-threshold is seen.


The spectrum predicted for the $1^+$ states is very similar to the
spectrum of the $0^+$ states. Fig. \ref{fig:1} (right panel) demonstrates that it
is shifted up by approximatively 140 MeV with respect to the $0^+$
spectrum. The bound state in the $(0,1)$-sector comes at 2440~MeV.
Thus the mass splitting of the $1^+$ and $0^+$ states in this
channel agrees very well with the empirical value of about 140~MeV
measured by the BABAR and CLEO collaborations \cite{BaBar,CLEO}. A
narrow structure at 2552~MeV is predicted in the
$(\frac{1}{2},0)$-channel which may be identified with the
$D(2420)$-resonance \cite{PDG02}. Even though the resonance mass
is overestimated by about 130~MeV our result is consistent with
its small width of about 20~MeV. The triplet state in this sector
of mass 2325~MeV has again a quite large width reflecting the
strong coupling to the $\pi\,D(2008) $-channel. Finally we obtain
strong cusp effects at the $\bar K \,D(2008)$- and  $K
\,D(2008)$-thresholds in the $(0,-1)$- and $(1,+1)$-sectors. It is
interesting to speculate whether chiral correction terms conspire
to slightly increase the net attraction in these sectors. This
would lead to a $(0,-1)$-bound state. The fact that we
overestimate the mass of the sextet state $D(2420)$ by about 130
MeV we take as a prediction that this should indeed be the
case \cite{HL03}. An analogous statement holds for the $0^+$
sector since due to heavy-quark symmetry chiral correction effects
in the $0^+$ and $1^+$ are identical at leading order.

\underline{To summarize:} We presented  a
coupled-channel description of the meson-meson scattering in the
open charm  sector using the chiral SU(3)
Lagrangian involving light-heavy $J^P=0^-$ and $J^P=1^-$ fields
that transform non-linearly under the chiral SU(3) group.
The major result of our study
is the prediction of the charmed mesonic states with $J^P=0^+,1^+$
quantum numbers forming anti-triplet and sextet representations of
the SU(3) group. This differs from the results implied by the
chiral quark model  leading to anti-triplet states only. Our
result suggests the existence of $J^P=0^+, 1^+$ states with
unconventional quantum numbers $(I,S)=(1,1)$ and $(I,S)=(0,-1)$.

\end{document}